\documentclass[fleqn,usenatbib]{mnras}

\usepackage{newtxtext,newtxmath}
\usepackage{url} 
\usepackage[T1]{fontenc}
\usepackage{subcaption}
\usepackage{orcidlink}
\usepackage{balance}
\usepackage{bm}

\DeclareRobustCommand{\VAN}[3]{#2}
\let\VANthebibliography\thebibliography
\def\thebibliography{\DeclareRobustCommand{\VAN}[3]{##3}\VANthebibliography}

\usepackage{graphicx}	% Including figure files
\usepackage{amsmath}	% Advanced maths commands

\title[{\sc MomentEmu}: A Generic Polynomial Emulator]{A general polynomial emulator for cosmology via moment projection}

\author[Z. Zhang]{
Zheng Zhang$^{1}$\orcidlink{0000-0002-9154-2803}\thanks{E-mail: zheng.zhang@manchester.ac.uk }
\\
$^{1}$Jodrell Bank Centre for Astrophysics, University of Manchester, Manchester, M13 9PL, UK
}

\date{Accepted XXX. Received YYY; in original form ZZZ}

\pubyear{\the\year{}}

\begin{document}
\label{firstpage}
\pagerange{\pageref{firstpage}--\pageref{lastpage}}
\maketitle

% Abstract of the paper
\begin{abstract} 
We present {\sc MomentEmu}, a general-purpose polynomial emulator for fast and interpretable mappings between theoretical parameters and observational features. The method constructs moment matrices to project simulation data onto polynomial bases, yielding symbolic expressions that approximate the target mapping. Compared to neural-network-based emulators, {\sc MomentEmu} offers negligible training cost, millisecond-level evaluation, and transparent functional forms.
As a proof-of-concept demonstration, we develop two emulators: {\sc PolyCAMB-$D_\ell$}, which maps six cosmological parameters to the CMB power spectra (TT, EE, BB, TE), and {\sc PolyCAMB-peak}, which enables a bidirectional mapping between the cosmological parameters and the acoustic peak features of $D_\ell^{\textsc{TT}}$. {\sc PolyCAMB-$D_\ell$} achieves sub-percent accuracy over multipoles $\ell \leq 4050$, while {\sc PolyCAMB-peak} also attains comparable precision and produces symbolic forms consistent with known analytical approximations.
The method is well suited for forward modelling, parameter inference, and uncertainty propagation, particularly when the parameter space is moderate in dimensionality and the mapping is smooth. {\sc MomentEmu} offers a lightweight and portable alternative to regression-based or black-box emulators in cosmological analysis.
\end{abstract}

% Select between one and six entries from the list of approved keywords.
% Don't make up new ones.

%%%%%%%%%%%%%%%%%%%%%%%%%%%%%%%%%%%%%%%%%%%%%%%%%%

%%%%%%%%%%%%%%%%% BODY OF PAPER %%%%%%%%%%%%%%%%%%

%%%%%%%%%%%%%%%%%%%%%%%%%%%%%%%
\section{Introduction}
\label{sec: introduction}
%%%%%%%%%%%%%%%%%%%%%%%%%%%%%%%

Cosmological parameter estimation increasingly relies on the use of fast surrogate models -- known as \emph{emulators} -- to replace expensive theoretical computations. A prominent example is the mapping between cosmological parameters and the Cosmic Microwave Background (CMB) angular power spectrum, traditionally evaluated by Boltzmann solvers such as {\sc CAMB} \citep{Lewis:1999bs} and {\sc CLASS} \citep{blas2011cosmic}. While numerically accurate, these solvers are slow for large-scale inference frameworks such as Markov Chain Monte Carlo (MCMC) or Approximate Bayesian Computation (ABC) \citep{cranmer2020frontier}.  

To address this, a wide range of emulators have been developed. These include neural network approaches \citep[e.g.,][]{auld2007fast,Cosmopower2022}, 
Gaussian-process regression \citep[e.g.,][]{lawrence2017mira},
polynomial regression \citep[e.g.,][]{fendt2007pico} and polynomial chaos \citep[e.g.][]{lucca2024crrfast}, symbolic regression methods \citep[e.g.,][]{bartlett2024precise}, and methods based on principal
component analysis (PCA) \citep[e.g.,][]{kwan2015cosmic}. 
Among these, neural emulators offer high performance, albeit at the expense of interpretability. 
In contrast, symbolic approaches are more transparent, but can be harder to scale due to expression depth increase and combinatorial growth in candidate expressions as the number of variables. Furthermore, regression-based methods tend to lack the flexibility required for retraining or incremental updates.

In this work, we present {\sc MomentEmu}\footnote{\href{https://github.com/zzhang0123/MomentEmu}{https://github.com/zzhang0123/MomentEmu}}, a simple, generic, and interpretable emulator based on moment projections and multivariate polynomial fits. Compared to regression-based polynomial methods, our approach avoids iterative fitting and instead constructs closed-form symbolic expressions via linear algebra on moment matrices. 
This allows both forward emulation (predicting observables from theory parameters) and backward emulation (inferring parameters from measured observables), with negligible numerical cost. The symbolic nature of the emulator makes it suitable for rapid error propagation, observable design, and interpretability-sensitive tasks such as emulator diagnosis and degeneracy exploration.

To demonstrate the power of {\sc MomentEmu}, we construct two emulators: {\sc PolyCAMB-$D_\ell$}, a fast surrogate for the CMB temperature power spectrum, and {\sc PolyCAMB-peak}, a bidirectional emulator for acoustic-peak features. Using a training set generated by {\sc CAMB}, we show that {\sc MomentEmu} achieves sub-percent accuracy at a second-level training speed and a millisecond-level full-spectrum evaluation speed,\footnote{On a Mac equipped with an Apple M3 Ultra chip. Similar equipment setup for other {\sc MomentEmu} runtime measurements apply and will not be repeated hereafter.} while preserving a high degree of symbolic transparency.  

The rest of the paper is organised as follows. In Section~\ref{sec: method} we present the methodology of {\sc MomentEmu}. In Section~\ref{sec: CMB applications} we apply it to CMB emulation: first to the temperature power spectrum (Section~\ref{sec: CMB ps}), and then to the acoustic-peak locations and amplitudes (Section~\ref{sec: CMB peaks}). We summarise and discuss implications in Section~\ref{sec: conclusion}.

\section{Method}
\label{sec: method}

\begin{figure*}
\begin{subfigure}{\textwidth}
  \centering
  \includegraphics[width=\textwidth]{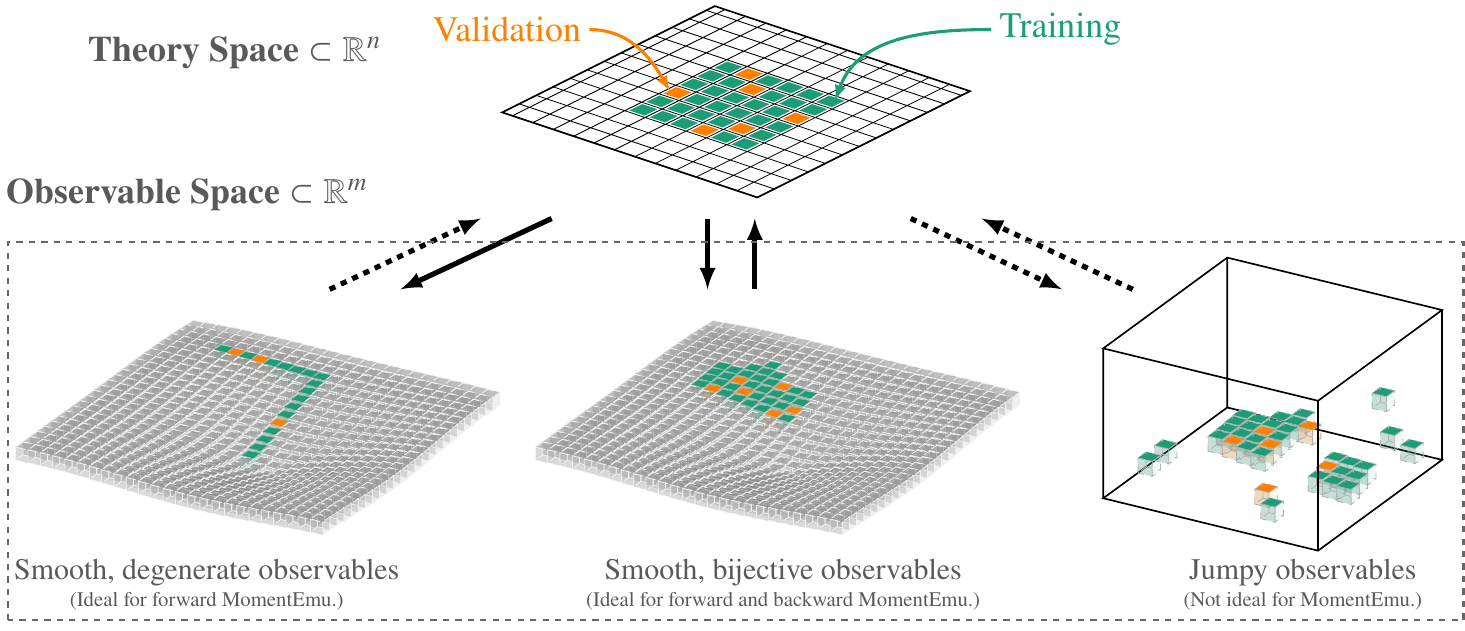}
  \caption{Mapping diagram: Conceptual illustration of the mappings between theory and observables. Solid arrows represent mappings that are well-suited to polynomial emulation, while dashed arrows indicate those that are less amenable to this approach. 
  Ideally, the test set should consist of random samples drawn independently from the parameter space. One should avoid constructing the training and test sets as disjoint subsets of the same regular grid, since in that case the test set cannot reveal potential overfitting of the emulator.}
  \label{fig: mapping diagram}
\end{subfigure}%

\begin{subfigure}{\textwidth}
  \centering
  \includegraphics[width=\textwidth]{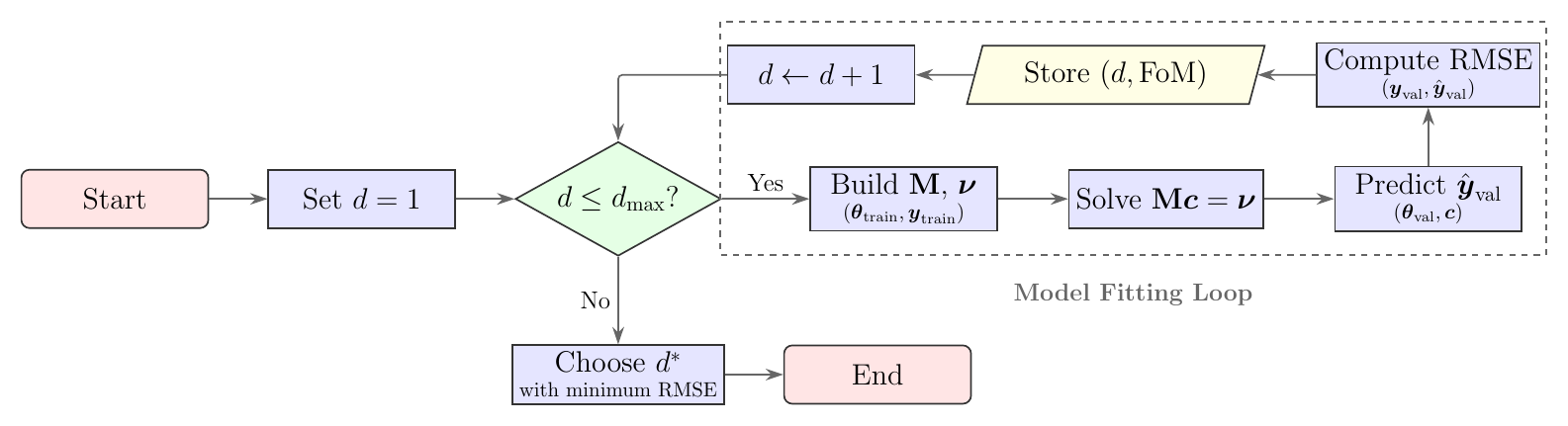}
  \caption{Workflow diagram: The full {\sc MomentEmu} workflow, as detailed in Section~\ref{sec: method}. Data and parameters are standardised for a stable numerical performance.}
  \label{fig: workflow diagram}
\end{subfigure}
\caption{ Diagrams illustrating how {\sc MomentEmu} operates. }
\label{fig: diagrams}
\end{figure*}

Let $\boldsymbol{\theta} \in \mathbb{R}^n$ denote theory parameters and $\boldsymbol{y} = \boldsymbol{y}(\boldsymbol{\theta})\in \mathbb{R}^m$ a set of scalar observables obtained as the ground-truth simulations. We approximate the forward model (i.e., the mapping from theory to observation) of each scalar observable by
\begin{equation}
\label{eq: poly approx}
\hat{y}_\ell(\boldsymbol{\theta}) = \sum_{\alpha \in \mathcal{A}_d} c_{\alpha\ell} \boldsymbol{\theta}^\alpha,
\end{equation}
where $\alpha=(\alpha_1, \ldots, \alpha_n)$ is a multi-index, and $\boldsymbol{\theta}^\alpha = \prod_{i=1}^n \theta_i^{\alpha_i}$ is a monomial.\footnote{For example, 
$y_1 = \theta_1^a + 2\theta_2^b + 3\theta_1^a\theta_2^b$ is denoted as $y_1 = \boldsymbol{\theta}^{\alpha_1} + 2\boldsymbol{\theta}^{\alpha_2} + 3\boldsymbol{\theta}^{\alpha_3}$ with $\alpha_1=(a, 0)$, $\alpha_2=(0,b)$ and $\alpha_3=(a, b)$.}
This equation generally represent a multivariate polynomial of degree (or order) $d$, as a linear combination of elements in
$ \mathcal{A}_d = \left\{ {\alpha} \in \mathbb{N}^n : |{\alpha}| \equiv \sum_{j=1}^n \alpha_j \leq d \right\} $ with $c_{\alpha\ell}$ the corresponding coefficients.

Given simulation data $\{(\boldsymbol{\theta}^{(i)}, \boldsymbol{y}^{(i)})\}_{i=1}^N$ with $i$ indexes the data points, we compute the moment matrix
%%%%%%%%%%%%%%%%%%%%%%%%%%%%%%%
\begin{equation}
\label{eq: mom mat}
M_{\alpha\beta}  = \frac{1}{N} \sum_{i=1}^N \phi_\alpha(\boldsymbol{\theta}^{(i)}) \phi_\beta(\boldsymbol{\theta}^{(i)}),
\end{equation}
%%%%%%%%%%%%%%%%%%%%%%%%%%%%%%%
where, for convenience, we have defined the monomial basis functions:
\(
\phi_{\alpha}(\boldsymbol{\theta}^{(i)}) = [\boldsymbol{\theta}^{(i)}]^{\alpha}.
\)
We also seek to obtain the projected targets (or moment vector):
%%%%%%%%%%%%%%%%%%%%%%%%%%%%%%%
\begin{equation}
\label{eq: mom vec}
\nu_{\alpha\ell} = \frac{1}{N} \sum_{i=1}^N y^{(i)}_\ell \phi_\alpha(\boldsymbol{\theta}^{(i)}).
\end{equation}
%%%%%%%%%%%%%%%%%%%%%%%%%%%%%%%
Under the assumption that the theory-to-observable mapping can be well-approximated by a multivariate polynomial, substituting Eq.~(\ref{eq: poly approx}) into Eq.~(\ref{eq: mom vec}) (replacing $y$) generates the linear system\footnote{Although the author initially derived this result using the moment-projection method, he subsequently recognized that an equivalent equation appears in the appendix of PICO, where it was obtained by imposing the vanishing of the first derivative of the chi-square.}
%%%%%%%%%%%%%%%%%%%%%%%%%%%%%%%
\begin{equation}
\label{eq: mom vec i.t.o. M c}
    \nu_{\alpha\ell} = \sum_{\beta=1}^D 
    c_{\beta\ell} M_{\alpha\beta} 
\end{equation}
%%%%%%%%%%%%%%%%%%%%%%%%%%%%%%%
where 
\begin{equation}
\label{eq: monomial basis dim}
    D=\vert\mathcal{A}_d \vert = 
    \frac{(n+d)!}{n! \, d!}
\end{equation} 
is the dimension of the monomial basis.
The solution of this system provides the linear coefficients $c_{\alpha\ell}$.
Equations~(\ref{eq: mom mat})–(\ref{eq: mom vec i.t.o. M c}) comprise the main numerical steps of {\sc MomentEmu}, highlighting the lightweight nature of the code. The algorithm is designed for the regime with many more training data than the monomial basis ($N\gg D$), in which the moment matrix (Eq.~(\ref{eq: mom mat})) is effectively guaranteed to be positive-definite. 

In practice, the optimal polynomial order $d$ is not known a priori. 
In the continuous domain, a higher-order emulator should, in principle, always outperform a lower-order one. However, with discrete and finite training data, overfitting can occur beyond a certain polynomial degree, leading to a turning point even when evaluated by RMSE alone. It is therefore crucial to use a representative test set -- e.g., one obtained by random sampling across the parameter space -- to ensure a fair assessment of predictive performance.
To address this, we implement a loop over $d$, starting from an initial guess and increasing up to a maximum degree specified by the user. This procedure selects either the best-fitting model or the first one that meets a predefined accuracy threshold. To protect against overfitting, the full set of simulations is partitioned into disjoint ``training'' and ``validation'' subsets. The training set is used to compute the polynomial coefficients for a given $d$, while the validation set is used to evaluate the root-mean-squared error (RMSE) of the standardised data, which serves as the figure of merit (FoM) for model selection. 
In addition to RMSE, \textsc{MomentEmu} supports auxiliary FoMs to quantify model complexity: among fits within a tolerance of the minimum RMSE, it selects the one with the smallest Bayesian Information Criterion (BIC) for compactness (see Appendix~\ref{app:model_selection} for details). Using Singular Value Decomposition (SVD), MomentEmu can also reduce dimensionality by thresholding the singular values of the functional-basis matrix.
For improved numerical stability, all parameters and observables are standardised (mean-centred and scaled by standard deviation) prior to fitting and transformed back to their original scales after the loop concludes. Figure~\ref{fig: diagrams} summarises the main steps of the {\sc MomentEmu} workflow.

The above procedure outlines how \textsc{MomentEmu} performs polynomial emulation of the forward mapping from theory parameters to observables. \textsc{MomentEmu} also supports the backward emulation, from observables back to theory parameters, by simply exchanging the roles of the input and output spaces. 
In order to construct a well-behaved inverse mapping, it is advisable to select a set of observables that will produce a smooth, continuous and non-degenerate transformation. Otherwise, one would need to resort to root-finding or algebraic geometric techniques to study the inverse mapping, both of which are considerably more complex than direct polynomial emulation.
We refer to the forward mapping as `observable prediction' and the inverse mapping as `parameter inference' to distinguish between these two operational modes.
%

%%%%%%%%%%%%%%%%%%%%%%%%%%%%%%%
\section{Application to CMB: PolyCAMB Emulators}
\label{sec: CMB applications}
%%%%%%%%%%%%%%%%%%%%%%%%%%%%%%%

%%%%%%%%%%%%%%%%%%%%%%%%%%%%%%%
\begin{table}
\centering
\caption{Parameter ranges used for generating training data with CAMB.}
\begin{tabular}{p{1.8cm}p{2cm}p{2cm}p{1cm}}
\hline
\textbf{Parameter} & \textbf{Range} & \textbf{Planck Best Fit} \\
\hline
  $\Omega_b h^2$  & [0.019, 0.025] 
  & $ 0.02242$\\

 $\Omega_c h^2$ & $[0.09, 0.15]$  
  & $ 0.11933$\ \\

 % $H_0$ [km/s/Mpc] & $[55.0, 80.0]$
 %  & $67.66$ \\
 
 $100\theta_\ast$  & $[1.00, 1.08]$ & 1.041\\
  
 $n_s$ & $[0.88, 1.02]$ 
  &  $ 0.9665$ \\

  $\ln(10^{10} A_s) $ & $ [2.70, 3.20] $ 
  & $3.047$  \\

  $\tau $ & $ [0.02, 0.12] $  
  & 0.0561 \\
\hline
\end{tabular}
\label{tab:params}
\end{table}
%%%%%%%%%%%%%%%%%%%%%%%%%%%%%%%
In this section, we apply {\sc MomentEmu} to CMB observables as a proof of concept, both to validate the method and to explore its key properties.
%%%%%%%%%%%%%%%%%%%%%%%%%%%%%%%

\subsection{Power Spectrum Emulator: {\sc PolyCAMB-\texorpdfstring{$D_\ell$}{Dl}} }
\label{sec: CMB ps}
%%%%%%%%%%%%%%%%%%%%%%%%%%%%%%%

%%%%%%%%%%%%%%%%%%%%%%%%%%%%%%%
\begin{figure}
    \centering
    \includegraphics[width=\linewidth]{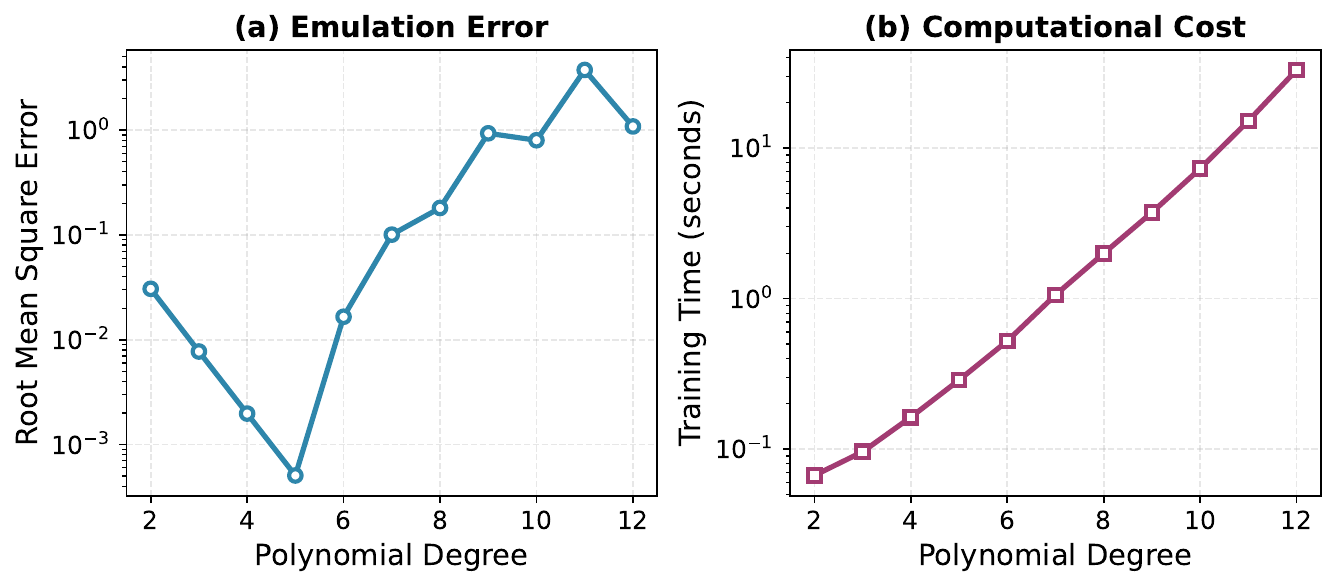}
    \caption{ 
    Emulator training performance (\textsc{PolyCAMB-$D_\ell^{\textsc{TT}}$}) vs polynomial degree. 
    }
    \label{fig: performance}
\end{figure}
%%%%%%%%%%%%%%%%%%%%%%%%%%%%%%%

We first apply it to CMB temperature power spectra. Specifically, we use the Boltzmann solver {\sc CAMB} to generate a set of 46,656 ($=6^6$) simulations on a regular grid, sampling the $6$-parameter flat 
$\Lambda$CDM model:
\begin{equation}
    \boldsymbol{\theta} =
    \left(\Omega_b h^2, \Omega_c h^2, \theta_\ast, n_s, \ln(10^{10}A_s), \tau \right)
\end{equation}
with parameter ranges listed in Table~\ref{tab:params}.
Each simulation maps theory parameters to the angular power spectrum, 
$D_\ell = {\ell(\ell+1)C_\ell}/{(2\pi)}$, computed for the TT, EE, BB, and TE components over the range $2\leq\ell\leq 4050$. We refer to this emulator as {\sc PolyCAMB-$D_\ell$}.

To illustrate the training performance, Figure~\ref{fig: performance} shows the RMSE and training time as functions of the polynomial degree. With a polynomial degree of $d=5$, {\sc PolyCAMB–$D_\ell$} achieves sub-percent accuracy: the standardised RMSE is about $0.05\%$ across the full multipole range, and the maximum deviation is $0.3\%$. All other emulators reach comparable sub-percent accuracy (with RMSE $\lesssim 0.1$\%); we therefore omit their detailed results here and refer readers to the {\sc PolyCAMB} documentation for a summary of their performance.
Typically, the emulator evaluation  takes $1.5$~ms per full $\ell$-range sample.\footnote{In general, running time scales with the degree of the polynomial and the number of $\ell$'s to be evaluated. Evaluating a list of parameter vectors together can further reduce per-sample evaluation time significantly.}
%
%%%%%%%%%%%%%%%%%%%%%%%%%%%%%%%
\begin{figure}
    \centering
    \includegraphics[width=\linewidth]{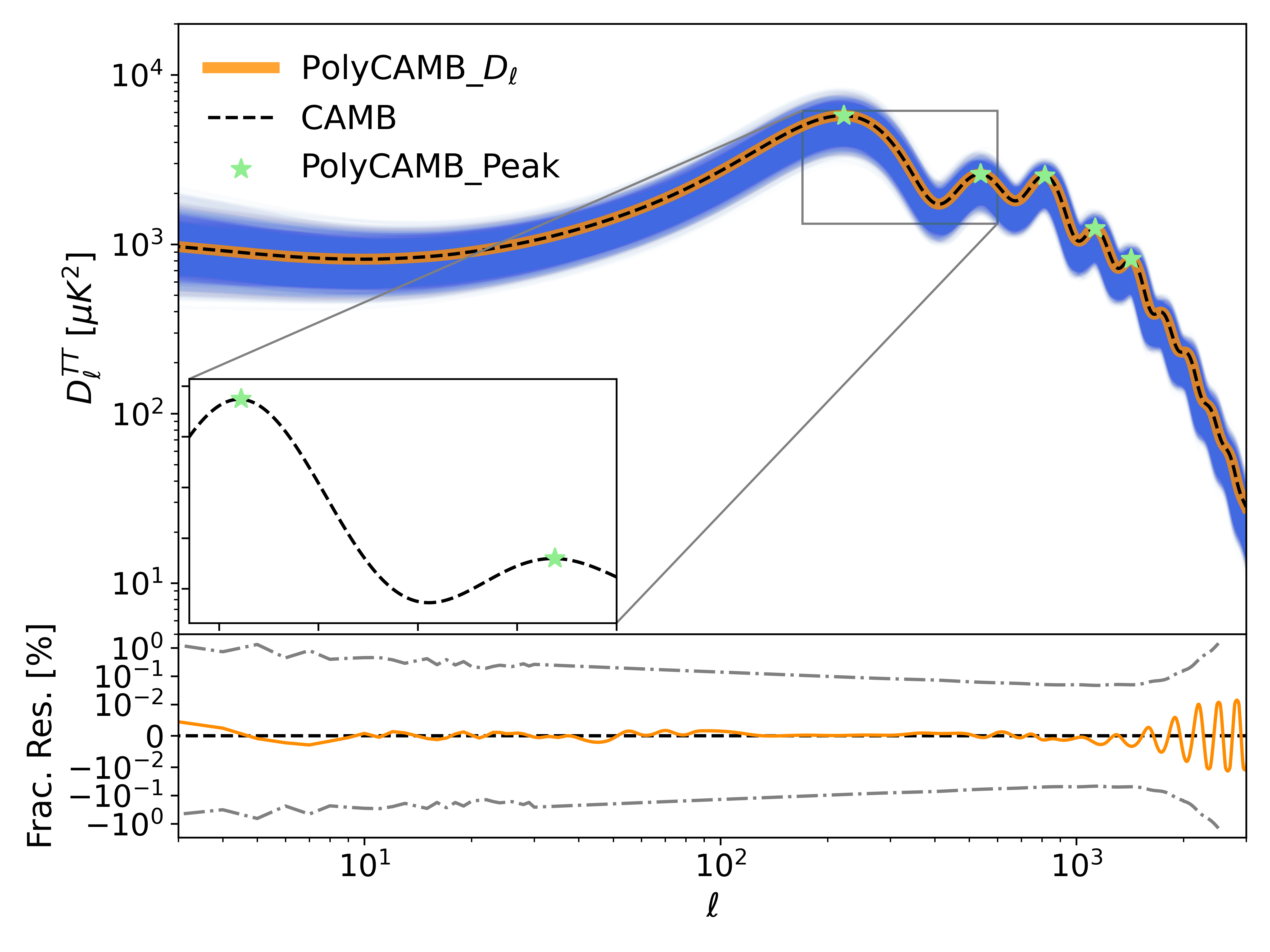}
    \caption{Validation of {\sc MomentEmu} with CMB observables.  
    (a) Comparison of $D_\ell^{\textsc{TT}}$ (top): the {\sc CAMB} spectrum (dashed line), and the {\sc PolyCAMB-$D_\ell$} emulation (thick orange). 
    The five star markers indicate the first five acoustic peaks as predicted by {\sc PolyCAMB-peak}.  
    The broad feature is an ensemble of emulator outputs (thin blue lines) generated from Gaussian perturbations of the input parameters, which illustrates a typical use case of fast forward modelling for Bayesian inference.
    (b) Fractional residuals (bottom): fractional differences between {\sc PolyCAMB-$D_\ell$} and {\sc CAMB}, with errors remaining below 0.02\% across the full multipole range. The gray dot–dashed lines indicate the Planck 68\% confidence interval (upper and lower error bars).
    }
    \label{fig: validation}
\end{figure}
%%%%%%%%%%%%%%%%%%%%%%%%%%%%%%%
%
Figure~\ref{fig: validation} compares the {\sc CAMB} spectrum with the {\sc PolyCAMB-$D_\ell$} prediction for a pivot cosmology\footnote{This model was outside the training set.} [chosen as the Planck best-$\Lambda$CDM \citep[the ``TT,TE,EE+lowE+lensing+BAO'' result in][]{aghanim2020planck}; summarised in Table~\ref{tab:params}], showing excellent agreement with a maximum fractional error below $0.02$\%.

{
To demonstrate this capability in a realistic inference setting, we use the {\sc PolyCAMB-$D_\ell$} emulator as a surrogate theory model within a full cosmological MCMC, combined with the \textit{Planck}~2018 ``TT,TE,EE+lowT+lowE'' likelihood. 
Sampling is performed using {\sc cobaya} \citep{torrado2021cobaya} with standard settings and flat priors for all the cosmological parameters.
See Appendix~\ref{append: mcmc setup} for more details on the MCMC setup.
}

%%%%%%%%%%%%%%%%%%%%%%%%%%%%%%%
\begin{figure*}
    \centering
    \includegraphics[width=0.85\linewidth]{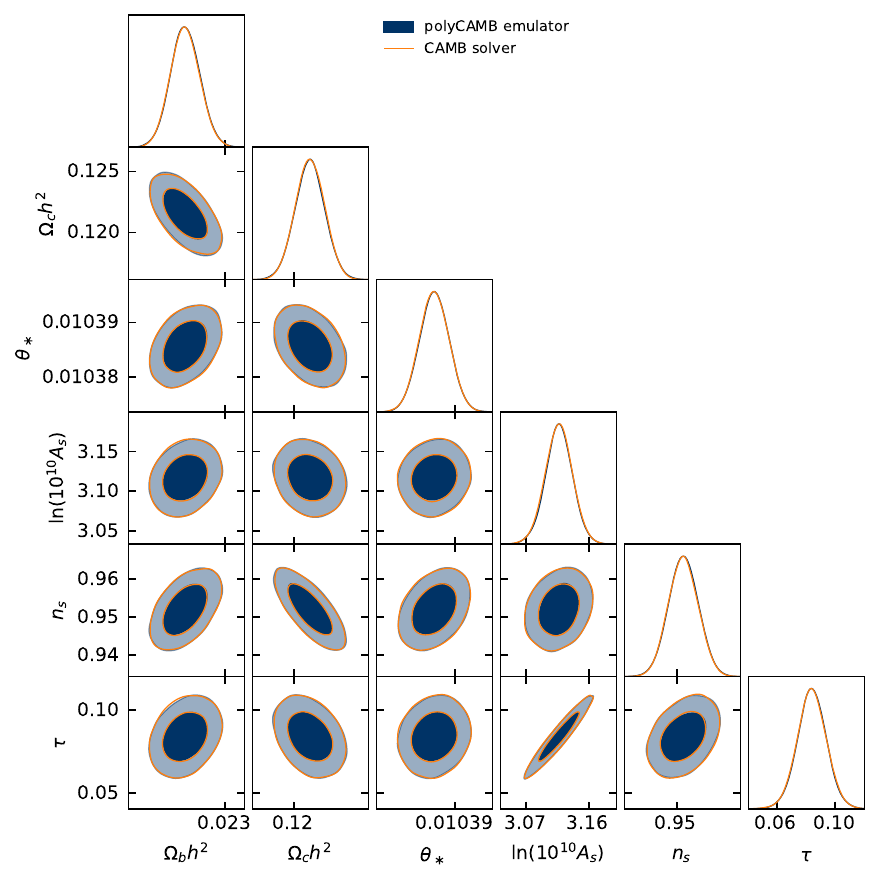}
    \caption{Corner plot showing the 68\% and 95\% joint posterior contours for the six $\Lambda$CDM parameters, derived from the Planck ``TT/TE/EE+lowE+lowT'' likelihoods using the raw \textsc{CAMB} and the \textsc{PolyCAMB-$D_\ell$} emulator separately, under identical MCMC sampling settings.
    One-dimensional marginalised posterior distributions are displayed along the diagonal panels, while the off-diagonal panels show the corresponding two-dimensional joint constraints. 
    We employed adaptive MCMC sampling with convergence determined by both sample count ($\geq200,000$ samples) and the Gelman-Rubin diagnostic ($R-1 < 0.02$), with additional confidence level monitoring ($R-1 < 0.2$ at 95\% confidence level) to ensure chain mixing and statistical reliability.
    Both corner plots are based on the last 200,000 samples to ensure comparable statistical robustness.
    The two contour sets exhibit good agreement, with minor discrepancies attributable to emulation errors, numerical differences, and sampling noise.
    All the best-fit parameters differs $\leq 0.01\sigma$ between using \texttt{polyCAMB} and \texttt{CAMB}.  
    }
    \label{fig: corner planck}
\end{figure*}
%%%%%%%%%%%%%%%%%%%%%%%%%%%%%%%
Figure~\ref{fig: corner planck} shows the posteriors for the six baseline $\Lambda$CDM parameters 
obtained after $2.8\times 10^5$ accepted MCMC steps (corresponding to $\sim 25$ minutes wall-clock time using 8 MPI ranks). The contours exhibit the expected degeneracy structures: while $A_s$ and $\tau$ are tightly coupled through their joint impact on the characteristic $A_s e^{-2\tau}$ amplitude, while the inclusion of low-$\ell$ polarisation data provides the main constraint on $\tau$. 
All recovered parameters are in excellent agreement with the corresponding \textsc{CAMB} results (which required $\sim$3530 minutes on the same hardware under identical MCMC settings). The shifts in the best-fit parameter values between \textsc{CAMB} and \textsc{polyCAMB} are all within $\lesssim0.01\sigma$. Within the accepted samples, the difference in the minimum $\chi^{2}$ between \textsc{polyCAMB} and \textsc{CAMB} is $0.2$, a statistically insignificant variation given that they correspond to slightly different points near the global maximum of the likelihood.
These results confirm the accuracy of {\sc MomentEmu} -- used here via {\sc PolyCAMB–$D_\ell$} for TT/TE/EE-based inference --as a reliable Boltzmann-solver surrogate, delivering an order-of-magnitude speed-up without compromising the integrity of the inferred posteriors.

We next discuss the performance and scalability of \texttt{PolyCAMB-$D_\ell$} for current and future CMB experiments. 
For a polynomial basis of degree $d$ with $N$ parameters, the number of basis functions is the combination 
$C_{N+d}^d = (N+d)!/(N!\,d!)$. 
To constrain a model in this basis, the training set must exceed this dimensionality. 
In the example of \textsc{PolyCAMB}, this is not a limitation: for \texttt{PolyCAMB-$D_\ell$} we find that a fifth-degree expansion provides an excellent fit, corresponding to roughly $500$ independent linear modes -- well below the size of our training data.

Regarding scalability, two aspects are relevant. 
First, increasing the number of cosmological parameters naturally enlarges the basis dimension, though our current setup has ample capacity to accommodate this. 
Second, as theoretical priors and data increasingly tighten parameter ranges, the required polynomial degree tends to decrease, partially offsetting the added complexity from extra parameters. 

Compared with neural-network approaches (such as \textsc{CosmoPower} \citep{Cosmopower2022}), a potential limitation of \texttt{MomentEmu} is the restricted parameter range: 
if the parameter space is too broad, the observable dependence may require a prohibitively high polynomial degree and potentially causing the emulation to break down. 
Fortunately, for CMB power spectra the dependence on cosmological parameters remains remarkably smooth over ranges far larger than their current uncertainties, making \texttt{MomentEmu} particularly well suited to this application.

%%%%%%%%%%%%%%%%%%%%%%%%%%%%%%%
\subsection{Acoustic Peak Emulator: {\sc PolyCAMB-peak}}
\label{sec: CMB peaks}
%%%%%%%%%%%%%%%%%%%%%%%%%%%%%%%

In addition to full power spectra, we also extract acoustic peak features as a compact summary of CMB observables. We use {\sc MomentEmu} to model both the forward and inverse mappings, i.e., from cosmological parameters to the locations and amplitudes of the first five acoustic peaks, and vice versa. The forward mapping is emulated with a polynomial degree of $2$ {at an accuracy level of $0.9$\%}, and the inverse mapping with degree $4$.
\footnote{We did not construct an inverse-mode emulator for {\sc PolyCAMB-$D_\ell$ }, as the high dimensionality of the observables ($4050$ 
$\ell$-modes) would require a significantly larger training set for stable inversion of moment matrix. While thinning the multipoles is possible, we consider the peak-feature-based inference more insightful and compact for parameter recovery.}

To facilitate discussion, we define:
%%%%%%%%%%%%%%%%%%%%%%%%%%%%%%%
\begin{itemize}
    \item $\ell_{p_k}$: the multipole location of the $k$-th peak
    \item $A_{p_k} = D^{\textsc{TT}}_{\ell_{p_k}}$: the corresponding peak amplitude/height
    \item $H_k = A_{p_k}/A_{p_1}$: relative peak heights
    \item $\eta_k = A_{p_k}/\ell_{p_k}$: scaled peak amplitudes
\end{itemize}
%%%%%%%%%%%%%%%%%%%%%%%%%%%%%%%
The set of observables used in this emulator is as follows:\footnote{In practice, we found that using the raw peak locations $\ell_{p_k}$ led to poor numerical performance. The alternative definition $\eta_k$, which retains positional information in a normalised form, resulted in significantly more stable behaviour.}
%%%%%%%%%%%%%%%%%%%%%%%%%%%%%%%
\begin{equation}
    \{\eta_1, \eta_2, \eta_3, \eta_4,\eta_5, 
    A_{p_1}, H_2, H_3, H_3, H_4\}.
\end{equation}
%%%%%%%%%%%%%%%%%%%%%%%%%%%%%%%
We denote this emulator as {\sc PolyCAMB-peak}. As shown in Figure~\ref{fig: validation}, the predicted peak positions and amplitudes for the pivot cosmology closely match those indicated directly by the temperature power spectrum.
%%%%%%%%%%%%%%%%%%%%%%%%%%%%%%%
\begin{figure*}
\begin{subfigure}{\textwidth}
  \centering
  \includegraphics[width=0.7\textwidth]{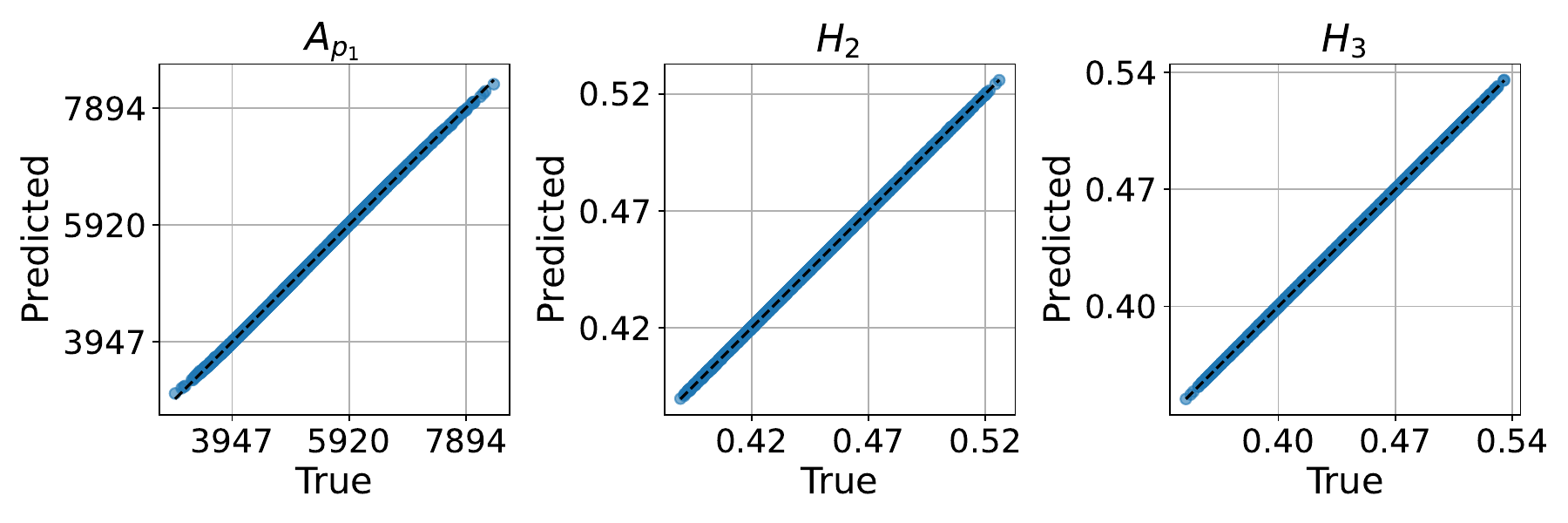}
  \caption{ \textbf{Observable prediction:} Comparison between predicted and true acoustic peak features using the forward mode of {\sc PolyCAMB-peak}. We display only the results for $\ln(10^{10}A_s)$, $H_2$, and $H_3$; the remaining observables show similarly close agreement with the true values and yield nearly identical plots, which we omit for brevity.}
  \label{fig: peak prediction part}
\end{subfigure}%

\begin{subfigure}{\textwidth}
  \centering
  \includegraphics[width=0.7\textwidth]{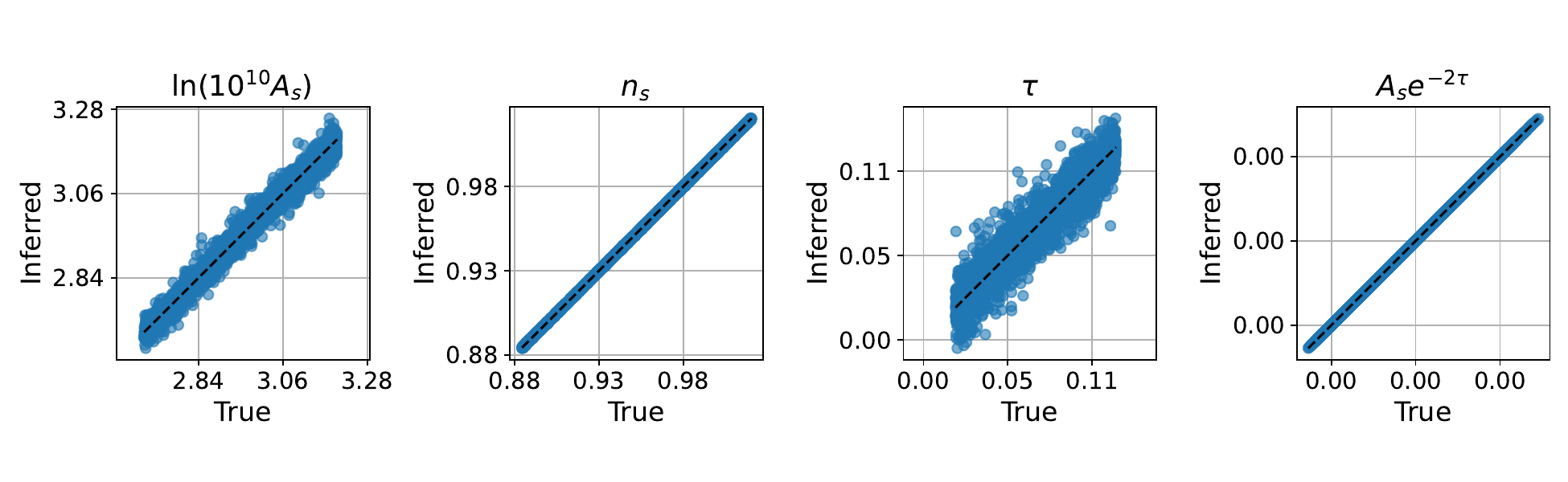}
  \caption{
  \textbf{Parameter inference:} Cosmological parameters recovered from acoustic peak features using the backward mode of {\sc PolyCAMB-peak}. The results show excellent agreement with the ground truth, except for $A_s$ and $\tau$, which remains weakly constrained due to its limited imprint on temperature peak structure alone. 
  Although $A_s$ and $\tau$ are individually poorly reconstructed due to their strong degeneracy, their combined amplitude $A_s e^{-2\tau}$ is well determined.
  The inferred values of $\Omega_bh^2$, $\Omega_ch^2$, and $\theta_\ast$ closely match their true inputs, yielding plots nearly identical to that of $n_s$; we omit these for conciseness. Note that this example assumes noiseless observables.
}
  \label{fig: peak inferrence}
\end{subfigure}
\caption{ Validation of bidirectional emulation using {\sc MomentEmu}.}
\label{fig: polycamb peak results}
\end{figure*}
%%%%%%%%%%%%%%%%%%%%%%%%%%%%%%%
Figure~\ref{fig: polycamb peak results} demonstrates both the observable prediction and parameter inference modes of {\sc PolyCAMB-peak}, evaluated on a held-out test set. As expected, predicted observables match their true values to high precision, and the inferred parameter values also show good agreement, with the notable exception of the optical depth $\tau$ and the magnitude  $A_s$. 
This is theoretically reasonable: the peak structure of the CMB temperature power spectrum carries little direct information about $\tau$, which primarily affects large-scale polarisation. Furthermore, $\tau$ is known to be degenerate with $A_s$, and this is reflected in a mild negative bias in the inferred values of $\ln(10^{10}A_s)$.
Thus, beyond accurate forward and inverse emulation, {\sc MomentEmu} also provides a physically interpretable framework for diagnostic analysis.

% Figure~\ref{fig:cl_example} shows the predicted vs. true spectrum and residuals. Figure~\ref{fig:performance} compares evaluation time and accuracy against neural emulators.

% \begin{figure}[ht!]
%     \centering
%     \includegraphics[width=\linewidth]{fig_perf_plot.pdf}
%     \caption{Left: Evaluation time vs. method. Right: emulator error vs. polynomial degree.}
%     \label{fig:performance}
% \end{figure}

%%%%%%%%%%%%%%%%%%%%%%%%%%%%%%%
\subsection{Symbolic Interpretability: Analytic Dependence of Peak Height}
\label{sec: symbolic peak height}
%%%%%%%%%%%%%%%%%%%%%%%%%%%%%%%
To further illustrate the symbolic nature and interpretability of {\sc MomentEmu}, we examine the closed-form polynomial expressions for the relative heights of the second and third acoustic peaks, $H_2$ and $H_3$, as produced by {\sc PolyCAMB-peak}. 
These observables are well-studied in the literature, notably by \citet{hu2001cosmic}, who provided approximate analytical formulae based on the physics of acoustic oscillations. In particular, the relative height of the second peak,
%%%%%%%%%%%%%%%%%%%%%%%%%%%%%%%
\begin{equation}
    \label{eq: H2 H01}
    H_2^{\text{(H01)}} =
    \frac{0.925 \,  \left(\omega_b + \omega_c\right)^{0.18} {2.4}^{n_s - 1}}{ \left[1+\left(\omega_b/0.0164\right)^{12 \left(\omega_b + \omega_c\right)^{0.52}} \right]^{1/5}},
\end{equation}
%%%%%%%%%%%%%%%%%%%%%%%%%%%%%%%
reflects the relative influence of baryon inertia (baryon loading) against photon pressure (radiation driving) in shaping the acoustic oscillations, while \citep{durrer2003acoustic}
\begin{equation}
    \label{eq: H3 H01}
    H_3^{\text{(H01)}} =
    \frac{2.17 \,  \left(\omega_b + \omega_c\right)^{0.59} {3.6}^{n_s - 1}}{ \left[1+\left(\omega_b/0.044\right)^{2} \right]
    \left[1+1.63(1-\omega_b/0.071)(\omega_b+\omega_c) \right]} 
\end{equation}
captures additional sensitivity to the matter density and damping scale.
For brevity, we have rewritten the density parameters as $\omega_b = \Omega_b h^2$ and $\omega_c = \Omega_c h^2$.

% $H_3$ is sensitive to the combined effects of matter density, which shapes gravitational potentials, and Silk damping, which suppresses small-scale fluctuations.

The expressions learnt by {\sc MomentEmu} also have a clear interpretation. Since the polynomial fit is constructed using mean-centred parameters, the resulting polynomial can be viewed as a truncated Taylor expansion\footnote{This is reminiscent of the moment expansion formalism investigated, for example, in \cite{chluba2017rethinking}.} of the observable around the mean of the parameter samples in the training set.
Although the coefficients may absorb contributions from regions far from the pivot\footnote{Taylor series capture the structures near the expansion's pivot better than those in regions far away, whereas a general polynomial fit doesn't overemphasise a particular region.} and higher-order terms due to truncation, we expect that the overall structure still captures the dominant smooth dependencies between parameters and observables.

To test this interpretation, we take the analytic expressions for $H_2$ and $H_3$ from \citet{hu2001cosmic} and perform a Taylor expansion about the mean cosmological parameters of our training set, up to the same polynomial degree ($d = 2$). For $H_2^{\text{(H01)}}$ we obtain
%%%%%%%%%%%%%%%%%%%%%%%%%%%%%%%
\begin{equation}
\begin{split}
    H_2^{\rm (H01)} =& \;
    161 \omega_b^{2} - 1.59 \omega_c^{2} + 0.176 n_s^{2} \\
    &- 77.3 \omega_b \omega_c - 12.2  \omega_b n_s + 0.215  \omega_c n_s \\
    &- 0.167 \omega_b  + 2.12 \omega_c + 0.311 n_s + 0.134
\end{split}
\end{equation}
%%%%%%%%%%%%%%%%%%%%%%%%%%%%%%%
The polynomial fit by {\sc PolyCAMB-peak} is
%%%%%%%%%%%%%%%%%%%%%%%%%%%%%%%
\begin{equation}
\begin{split}
    H_2^{\text{(Z25)}} = & \;
    175 \omega_b^{2} - 1.26 \omega_c^{2} + 0.161 n_s^{2} \\
    &- 46.2 \omega_b \omega_c - 9.73 \omega_b n_s + 0.257 \omega_c n_s \\
    &- 5.76 \omega_b + 1.18 \omega_c + 0.282 n_s + 0.179  \\
    & + \text{remaining terms}
\end{split}
\end{equation}
%%%%%%%%%%%%%%%%%%%%%%%%%%%%%%%
Similarly, the expanded $H_3^{\text{(H01)}}$ is
%%%%%%%%%%%%%%%%%%%%%%%%%%%%%%%
\begin{equation}
\begin{split}
    H_3^{\text{(H01)}} =& - 73.9 \omega_b^{2} - 4.03 \omega_c^{2} + 0.364 n_s^{2} \\
    & - 22.2 \omega_b \omega_c - 6.93  \omega_b n_s + 1.81  \omega_c n_s \\
    & + 7.09 \omega_b  + 1.15 \omega_c - 0.188 ns  + 0.0907
\end{split}
\end{equation}
%%%%%%%%%%%%%%%%%%%%%%%%%%%%%%%
and the counter part given by {\sc PolyCAMB-peak} is
%%%%%%%%%%%%%%%%%%%%%%%%%%%%%%%
\begin{equation}
\begin{split}
    H_3^{\text{(Z25)}} = & 
    - 82.7 \omega_b^{2} - 3.96 \omega_c^{2} + 0.298 n_s^{2}\\
    & - 18.9 \omega_b \omega_c - 6.52 \omega_b n_s  + 1.44 \omega_c n_s \\
    & + 6.26 \omega_b + 1.20 \omega_c - 0.0669 n_s -0.0375 \\
    & + \text{remaining terms}
\end{split}
\end{equation}
%%%%%%%%%%%%%%%%%%%%%%%%%%%%%%%
The symbolic emulators above offer a response-function view of the peak-height ratio sensitivity. Interpreted in this way, the coefficients in $H_2^{\text{(Z25)}}$ exhibit the expected signs from acoustic-physics arguments, revealing how each parameter drives the relative peak amplitudes. Several insights emerge from these trends, as reflected in both the symbolic expression and Figure~\ref{fig: H01 vs Z25}:
\begin{enumerate}
    \item $A_s$ or $\tau$ do not appear in these equations, since they only linearly scale the power spectrum, so they do not affect the ratio between peaks. 

    \item Increasing $\omega_b$ lowers $H_2$, as baryon loading enhances the contrast between compressional (odd) and rarefaction (even) modes, amplifying the first peak while suppressing the second. This produces the negative linear and mixed $\omega_b$-terms in the emulator; the positive $\omega_b^2$ term indicates saturation of this effect at higher $\omega_b$.

    \item Increasing the total matter density $\omega_m = \omega_b + \omega_c$ shifts matter-radiation equality to earlier epoch, shortening the time of potential decay that drives acoustic oscillations. This reduces the radiation driving of all modes, but affects the first peak more strongly than the second, since longer-wavelength modes entered the horizon earlier and depended more on that decay. The result is a net increase of $H_2$ with $\omega_c$. The small negative $\omega_c^2$ term in the emulator reflects the expected saturation of this effect once equality occurs well before recombination.

    \item A larger $n_s$ enhances small-scale primordial power, increasing the second-to-first peak ratio $H_2$. The negative $\omega_b n_s$ and positive $\omega_c n_s$ cross-terms reflect how this sensitivity depends on baryon loading and matter content: baryons damp the second peak and weaken the $n_s$-driven rise, while higher $\omega_c$ shifts equality earlier and strengthens it. 
\end{enumerate}
This illustrative TT-peak example demonstrates how {\sc MomentEmu} provides a response-function perspective on how observables respond to theoretical parameters. While here the trends mirror well-understood CMB acoustic physics, in more general applications {\sc MomentEmu} can reveal emergent response laws in regimes where no analytic theory exists, offering valuable physical insight into previously unexplored behaviours.

{Roughly speaking, the analytical approximations of $H_2$ and $H_3$ presented in \citet{hu2001cosmic} show good agreement with those produced by {\sc PolyCAMB-peak}, both in functional structure and leading-order parameter dependencies. Some deviations are expected, given that the analytical forms are designed primarily for qualitative insight \citep{hu2001cosmic}, and that our low-order polynomial fits are not guaranteed to exactly reproduce a Taylor expansion.
%
%%%%%%%%%%%%%%%%%%%%%%%%%%%%%%%
\begin{figure*}
    \centering
    \includegraphics[width=0.8\linewidth]{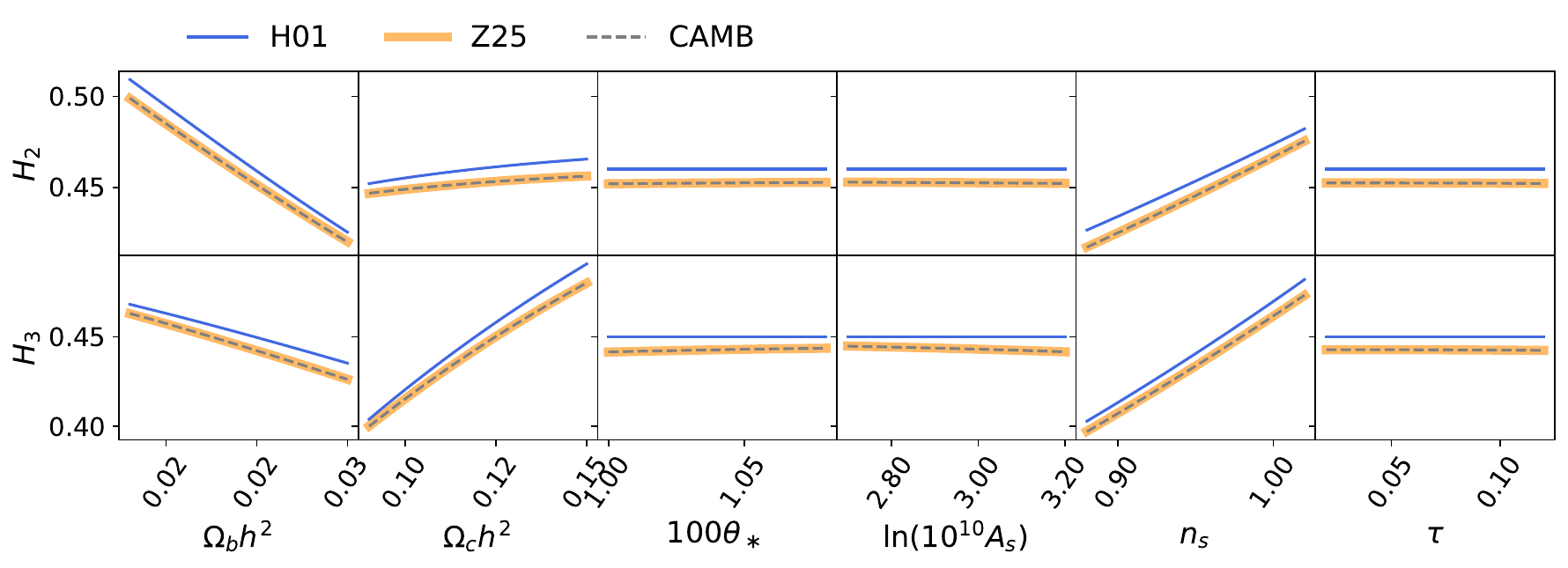}
    \caption{
    {
    Comparison of the $H_2$ and $H_3$ peak height ratios obtained from the analytical approximations of \citet{hu2001cosmic} (``H01''; thin blue), our polynomial emulator {\sc PolyCAMB-peak} (``Z25''; thick orange), and the true values from CAMB simulations (``CAMB''; dashed gray). 
    All curves are shown as functions of a single varying parameter, with the remaining cosmological parameters fixed at the pivot model.
    The overall trends and parameter sensitivities (primarily to $\Omega_b h^2$, $\Omega_c h^2$, and $n_s$) are consistent across all methods. 
    Amplitude differences remain modest: taking the CAMB results as reference, the accuracy is $\sim 0.04\%$ for the {\sc Z25} expressions, and  $1.7\%$ for $H_2$ and $1.6\%$ for $H_3$ in the H01 approximation -- the later is well within the $\sim$5\% accuracy quoted in \citet{durrer2003acoustic}.
    }
    }
    \label{fig: H01 vs Z25}
\end{figure*}
%%%%%%%%%%%%%%%%%%%%%%%%%%%%%%%
Figure~\ref{fig: H01 vs Z25} provides a quantitative comparison between the analytic approximations (H01; Eqns.~(\ref{eq: H2 H01}) and (\ref{eq: H3 H01})), our emulator (Z25), and the CAMB-fitted reference values. Despite modest amplitude differences, the overall trends and parameter sensitivities remain consistent, well within the expected accuracy range for such acoustic peak approximations.
}

Note that since the analytic approximations from \citet{hu2001cosmic} depend only on three parameters, while {\sc PolyCAMB-peak} fits all six cosmological parameters, for ease of comparison we retain only the monomial terms shared with \citet{hu2001cosmic}. The remaining terms, involving additional parameters, are considered subdominant. 
The full six-parameter, second-order polynomial emulations for $H_2$ and $H_3$ are presented in Appendix~\ref{append: full H2 H3}. 
Polynomial expressions for the other observables, as well as the inverse mappings of cosmological parameters as functions of acoustic peak observables, are available in the {\sc MomentEmu} GitHub repository notebooks {(See Data Availability for details)}.

This agreement underscores the \textit{symbolic transparency} of {\sc MomentEmu}: its output can be directly interpreted as a data-driven, low-order Taylor expansion of established physical relationships.

These symbolic expressions provide explicit and interpretable mappings between cosmological parameters and acoustic peak features, facilitating semi-analytic sensitivity analyses, tracing of parameter dependencies, and the construction of compact surrogate models for theory-to-observable mappings.

% \paragraph*{Peak location and height.} 

% The first acoustic peak in the CMB temperature power spectrum encodes key information about the geometry and contents of the Universe. While the full $D_\ell$ spectrum is obtained by solving the Boltzmann equations numerically, approximate analytic expressions capture the leading dependence of the peak’s position and amplitude on the six $\Lambda$CDM parameters.

% The position $\ell_1$ of the first peak is approximately given by the angular scale of the sound horizon at recombination:
% \begin{equation}
% \ell_1 \approx \pi \frac{d_A(z_*)}{r_s(z_*)},
% \end{equation}
% where $d_A(z_*)$ is the comoving angular diameter distance to last scattering and $r_s(z_*)$ is the comoving sound horizon at that epoch. For flat $\Lambda$CDM, both quantities depend primarily on the physical densities $\Omega_b h^2$, $\Omega_c h^2$, and the Hubble parameter $H_0$.

% The height $H_1 \equiv D_{\ell_1}$ of the first peak is influenced by multiple effects, including baryon loading, the early Integrated Sachs–Wolfe (ISW) effect, and the primordial amplitude. Its approximate scaling is given by:
% \begin{equation}
% H_1 \propto A_s \, e^{-2\tau} \left(1 + \frac{\Omega_b h^2}{\Omega_c h^2} \right)^{2.5} \left( \Omega_b h^2 + \Omega_c h^2 \right)^{-0.5}.
% \end{equation}
% This expression captures the enhancement of compressional modes by baryons, suppression due to reionization, and the amplitude normalization from the primordial power spectrum.

%%%%%%%%%%%%%%%%%%%%%%%%%%%%%%%
\section{Discussion and conclusions}
\label{sec: conclusion}
%%%%%%%%%%%%%%%%%%%%%%%%%%%%%%%

We have introduced \textsc{MomentEmu}, a moment-based, general-purpose polynomial emulator for any smooth mapping between theory parameters and observables. To demonstrate its validity, negligible numerical cost, and high degree of interpretability, we produced two illustrative by-products: \textsc{PolyCAMB–$D_\ell$}, which emulates the CMB temperature power spectrum, and \textsc{PolyCAMB–peak}, which emulates the bidirectional mapping between cosmological parameters and acoustic peak features.  Below we summarise the key properties of \textsc{MomentEmu}.

\paragraph*{Speed‐up: inexpensive training and evaluation.}
In the common regime where the training-set size is much larger than the polynomial basis dimension ($N \gg D$), the dominant cost is assembling the moment matrix
(Equation~\ref{eq: mom mat}), which scales as $\mathcal{O}(N D^{2})$.  
For the moderate polynomial degrees typically required, this cost is modest, and can be reduced further by sampling parameters on a grid and caching intermediate monomial products.  Consequently, the overall complexity is comfortably below $\mathcal{O}(N D^{2})$.
For example, using an Apple M3 Ultra chip, \textsc{PolyCAMB–$D_\ell$} fits $6$ parameters to $2{,}510$ observables with a fifth-order polynomial, using $\sim\! 1.1 \times 10^{5}$ regular grid simulations, in $\sim\! 9$\,s -- orders of magnitude faster than a typical neural-network workflow such as \textsc{CosmoPower}\footnote{The training time of \textsc{CosmoPower} can be found in its accompanying Colab notebook [see the Data Availability section], which reports approximately $15$~minutes for a 5-step training cycle on a dataset of $\sim 5~\times 10^4$ simulations, using a Google Compute Engine GPU backend.}.  
Spectrum evaluation is equally fast: a full set of $D_\ell$ values is produced in $\sim\! 1$\,milliseconds.  Because both training \emph{and} inference are inexpensive, \textsc{MomentEmu} is ideal for iterative or rapid-turnaround analysis pipelines.

\paragraph*{Versatility, universality, and scalability.}
The same workflow applies unchanged to any smooth theory–observable map, from 21\,cm power spectra to large-scale-structure summaries.  The forward mode (observable prediction) is naturally suited to high-dimensional Bayesian inference, while the backward mode (parameter inference) provides a transparent surrogate for likelihood-free or simulator-based inference.  It also helps to design reduced but informative observables and diagnose parameter degeneracies, as illustrated by the low sensitivity of acoustic-peak data to the optical depth~$\tau$ in \textsc{PolyCAMB–peak}.  
Scaling with training-set size is linear, so larger data sets are easily accommodated.  
Although the $D^{2}$ term means cost can rise with many parameters or very high polynomial degree, most cosmological observables are sufficiently smooth that low orders suffice in large parameter spaces; if necessary, one can partition parameter space into several local patches.

\paragraph*{Interpretability.}
\textsc{MomentEmu} returns fully symbolic expressions for theory–observable relations.  Unlike neural network symbolic regressions, these polynomials are transparent; as shown in Section~\ref{sec: symbolic peak height}, they can be interpreted as truncated Taylor expansions about the mean of the training set.  We refer to this property as \emph{symbolic transparency}.  It enables analytic sensitivity calculations, closed-form derivatives, and straightforward physical insight.

\paragraph*{Differentiability.}
An important advantage of the moment-projection polynomial emulator is that the resulting symbolic expressions are fully differentiable with respect to input parameters. This property enables efficient and exact evaluation of derivatives, which is particularly valuable for applications such as Fisher matrix forecasts, gradient-based optimization, and sensitivity analyses. 

\paragraph*{Portability.}
MomentEmu produces highly compact polynomial emulators compared to their training datasets. 
For example, while the training data for \textsc{PolyCAMB-}$D_\ell$ occupies roughly 2\,GB, the resulting emulator file is about 33\,MB, and \textsc{PolyCAMB-peak} is an even smaller 0.05\,MB -- excluding the separately storable symbolic expressions. 
This reduction in size makes {\sc MomentEmu} models extremely portable and convenient to share or deploy in computational pipelines without significant data transfer overhead.

\paragraph*{Extensions.}
The formulation of \textsc{MomentEmu} can be extended in more general directions:
(1)
In this work, we project the data onto a set of basis functions and then recover the coefficients by inverting the moment matrix. In principle, one could generalize this by contracting the data with an order‑$n$ tensor and inverting a corresponding order‑$(n+1)$ tensorial moment structure.
(2)
We have used a polynomial basis, which allows the resulting fit to be interpreted as a truncated Taylor expansion when training over a small region. However, this choice is not essential: the framework is compatible with any complete and well-behaved function basis, not just polynomials.

\paragraph*{Limitations.}
First of all, like any emulator, \textsc{MomentEmu} relies on a high-fidelity training set -- in our example produced by \textsc{CAMB}.  Its accuracy also depends on the smoothness of the underlying mapping, as illustrated in Figure~\ref{fig: mapping diagram}. 
%only when the mapping is smooth can MomentEmu work well, 
%for most cosmological observables this is not limiting.  

Second, as the sampled parameter volume increases, the accuracy decreases and/or the polynomial degree increases. Therefore,  \textsc{MomentEmu} is best suited to problems where the region of interest is already roughly known. In contrast, neural networks such as \textsc{CosmoPower} can more easily cover a very wide range of parameters.  Users may therefore trade coverage for speed by shrinking the parameter domain or by fitting several local patches.

Third, \textsc{MomentEmu} does not guarantee accurate fits outside the training region. This limitation can be understood in two ways: as a truncated Taylor expansion and as a general issue inherent to polynomial fitting.

\medskip\noindent
In summary, \textsc{MomentEmu} offers a fast, interpretable and flexible alternative to black-box emulators. This makes it particularly attractive when rapid retraining or explicit symbolic forms are desirable.

\section*{Acknowledgements}
The author would like to thank Philip Bull and Jens Chluba for their helpful comments.
The reviewer's thoughtful and constructive feedback has greatly strengthened the manuscript and improved the demonstration of {\sc MomentEmu}, particularly in its application to CMB power spectra.
Thanks also to Davide Piras and Alessio Spurio Mancini for clarifying the training time for \textsc{CosmoPower}.
The results were obtained as part of a project that has received funding from the European Research Council (ERC) under the European Union's Horizon 2020 research and innovation programme (Grant agreement No. 948764).  
The author also acknowledges support from the RadioForegroundsPlus project HORIZON-CL4-2023-SPACE-01, GA 101135036.

%%%%%%%%%%%%%%%%%%%%%%%%%%%%%%%%%%%%%%%%%%%%%%%%%%
\section*{Data Availability}
All code, data, and Jupyter notebooks necessary to reproduce the results presented in this paper are available in the associated GitHub repository: \url{https://github.com/zzhang0123/MomentEmu}.
The CosmoPower training notebook referenced in Section~\ref{sec: conclusion} is available at:  
\url{https://colab.research.google.com/drive/1eiDX_P0fxcuxv530xr2iceaPbY4CA5pD?usp=sharing}.
The Planck error bar information is obtained from \cite{ESA2015_PHZ}: \url{https://esdcdoi.esac.esa.int/doi/html/data/astronomy/planck/Cosmology.html}.

%%%%%%%%%%%%%%%%%%%% REFERENCES %%%%%%%%%%%%%%%%%%

% The best way to enter references is to use BibTeX:

\bibliographystyle{mnras}
\bibliography{MomentEmu} % if your bibtex file is called example.bib

% Alternatively you could enter them by hand, like this:
% This method is tedious and prone to error if you have lots of references
%\begin{thebibliography}{99}
%\bibitem[\protect\citeauthoryear{Author}{2012}]{Author2012}
%Author A.~N., 2013, Journal of Improbable Astronomy, 1, 1
%\bibitem[\protect\citeauthoryear{Others}{2013}]{Others2013}
%Others S., 2012, Journal of Interesting Stuff, 17, 198
%\end{thebibliography}

%%%%%%%%%%%%%%%%%%%%%%%%%%%%%%%%%%%%%%%%%%%%%%%%%%

%%%%%%%%%%%%%%%%% APPENDICES %%%%%%%%%%%%%%%%%%%%%

\appendix

\section{Information Criteria on Test Data}
\label{app:model_selection}

Beyond the standard predictive metrics, RMSE, 
we assess model complexity using information-based criteria adapted to test data. 
For a model with $k$ free parameters and Gaussian errors, the Akaike and Bayesian Information Criteria are
\begin{align}
\mathrm{AIC} &= 2k - 2 \ln \hat L_\mathrm{train}, &
\mathrm{BIC} &= k \ln n_\mathrm{train} - 2 \ln \hat L_\mathrm{train},
\end{align}
where $\hat L_\mathrm{train}$ is the likelihood of the training data and 
$n_\mathrm{train}$ the number of training samples. 
To assess predictive performance, we define \emph{test-set} analogues by replacing 
the training likelihood with that of the test data. 
For a test set $\{y_i,\hat y_i\}$ of size $n_\mathrm{test}$, the mean squared error is 
$\mathrm{MSE} = n_\mathrm{test}^{-1}\!\sum_i (y_i - \hat y_i)^2$, and under Gaussian errors,
\begin{align}
\mathrm{AIC}_\mathrm{test} &= n_\mathrm{test}\ln(\mathrm{MSE}) + 2k, &
\mathrm{BIC}_\mathrm{test} &= n_\mathrm{train}\ln(\mathrm{MSE}) + k\ln n_\mathrm{train}.
\end{align}
If $n_\mathrm{train}$ is unavailable, $n_\mathrm{test}$ may be used heuristically. 
These predictive criteria jointly penalize poor fit and excessive complexity, 
providing an interpretable basis for model comparison.

For model selection, we adopt a two-step strategy. 
First, we filter by predictive accuracy: models with test-set RMSE within a tolerance 
of $\delta = 0.05$ of the minimum value,
\[
\mathrm{RMSE} \le \mathrm{RMSE}_{\min}(1+\delta),
\]
are retained as competitive candidates. 
Second, we choose the simplest of these by minimizing the information criterion 
(preferably BIC, or AIC otherwise), computed from the residual sum of squares (RSS) as
\[
\mathrm{AIC}=2k+n\ln(\mathrm{RSS}/n), \qquad
\mathrm{BIC}=k\ln n+n\ln(\mathrm{RSS}/n),
\]
where $n$ is the test-sample size. 
This procedure balances predictive performance and model parsimony, 
favoring models that generalize well without unnecessary complexity.

\section{MCMC Parameter Estimation Details}
\label{append: mcmc setup}

\subsection{Likelihood and Data}

We perform Bayesian parameter estimation using Planck 2018 data, incorporating three likelihood components:
\begin{itemize}
    \item \textbf{Low-$\ell$ TT}: Commander likelihood for temperature anisotropies at $\ell < 30$
    \item \textbf{Low-$\ell$ EE}: SimAll likelihood for E-mode polarization at $\ell < 30$  
    \item \textbf{High-$\ell$ TTTEEE}: Plik likelihood for temperature and polarization cross-correlations at $30 \leq \ell \leq 2508$
\end{itemize}

The theoretical power spectra are computed using the PolyCAMB emulator, which provides fast and accurate predictions for the angular power spectra across the full multipole range.

\subsection{Parameter Space and Priors}

We sample six standard $\Lambda$CDM cosmological parameters with uniform priors:
\begin{itemize}
    \item $\omega_b$: Baryon density parameter $[0.019, 0.025]$
    \item $\omega_c$: Cold dark matter density parameter $[0.09, 0.15]$  
    \item $\theta_*$: Angular scale of sound horizon $[0.0100, 0.0108]$
    \item $\ln(10^{10}A_s)$: Primordial scalar amplitude $[2.7, 3.2]$
    \item $n_s$: Spectral index $[0.88, 1.02]$
    \item $\tau$: Optical depth to reionization $[0.02, 0.12]$
\end{itemize}

Additionally, we include the Planck calibration nuisance parameter $A_{\text{Planck}}$ with a conservative flat prior $[0.9, 1.1]$.

\subsection{MCMC Configuration}
In the \textsc{cobaya} parameter estimation, we employ the Metropolis-Hastings sampler with the following optimizations:
\begin{itemize}
    \item \textbf{Proposal learning} enabled with $R-1$ threshold of 3.0 for adaptation
    \item \textbf{Convergence criteria}: $R-1 < 0.02$ for means and $R-1 < 0.2$ for 95\% confidence intervals
    \item \textbf{Dragging} enabled with limits $[0.05, 0.25]$ to improve sampling efficiency
    \item \textbf{Oversampling} with power $0.4$ and thinning for MPI parallelization
    \item \textbf{Proposal scale}: $2.4$ (standard for $6$-dimensional parameter space)
\end{itemize}

Convergence is monitored every $40 \times d$ samples (where $d$ is the dimensionality), with a maximum of 200,000 samples per chain. Progress is output every $5$ minutes to facilitate monitoring of long MPI runs.

The reference values for parameter initialization are taken from the Planck 2018 best-fit cosmology, with proposal widths tuned to account for known parameter degeneracies (particularly between $\ln(10^{10}A_s)$ and $\tau$).

% \paragraph{Convergence.}
% Based on the progress file, the MCMC chains show good convergence behavior. The Gelman-Rubin statistic $R-1$ decreases from $\sim 10$ initially to $< 0.02$ after approximately 280,000 samples, indicating excellent convergence. The acceptance rate stabilizes around 27\%, which is within the optimal range for efficient sampling. The confidence level statistic $R-1_{\text{cl}} = 0.096$ at the final checkpoint confirms that the 95\% confidence intervals are well-converged across all parameters.

\subsection{Convergence and Performance Comparison}

We performed two identical MCMC runs using the same Planck 2018 likelihoods and parameter configuration, but with different Boltzmann solvers: the standard CAMB code and the PolyCAMB emulator.

\subsubsection{PolyCAMB Performance}
The PolyCAMB emulator demonstrates excellent convergence behavior. The Gelman-Rubin statistic $R-1$ decreases rapidly from $\sim 10$ initially to $< 0.02$ after approximately 280,000 samples. The acceptance rate stabilizes around 27\%, which is within the optimal range for efficient sampling. The confidence level statistic $R-1_{\text{cl}} = 0.096$ at the final checkpoint confirms well-converged 95\% confidence intervals. The total runtime was approximately 23 minutes.

\subsubsection{CAMB Performance}
The standard CAMB run shows similar convergence quality but with significantly longer computation time. The $R-1$ statistic follows a comparable trajectory, reaching $< 0.02$ after about 245,000 samples. The acceptance rate stabilizes around 27\%, identical to the PolyCAMB run. The final $R-1_{\text{cl}} = 0.106$ indicates equally good convergence of confidence intervals. However, the total runtime extended to approximately 3500 minutes.

Both runs achieve identical statistical convergence quality, demonstrating that the PolyCAMB emulator maintains full accuracy while providing order-of-magnitude computational savings. This efficiency gain becomes particularly valuable for large-scale cosmological surveys and extensive parameter space exploration.

\section{Symbolic expressions for \texorpdfstring{$H_2^{(Z25)}$ and $H_3^{(Z25)}$}{H2(Z25) and H3(Z25)}}
\label{append: full H2 H3}

This appendix provides the full symbolic expressions for two key observables: the relative heights of the second ($H_2$) and third ($H_3$) acoustic peaks with respect to the first peak. These are emulated by {\sc PolyEmu\_peak} using second-order polynomial expansions in the six $\Lambda$CDM parameters.

The second-order polynomial expression for $H_2$ is given by
\begin{equation}
\begin{split}
    H_2^{\text{(Z25)}} = & \;
    175 \omega_b^{2} - 1.26 \omega_c^{2} + 0.161 n_s^{2} \\
    &- 46.2 \omega_b \omega_c - 9.73 \omega_b n_s + 0.257 \omega_c n_s \\
    &- 5.76 \omega_b + 1.18 \omega_c + 0.282 n_s + 0.179 \\
    &- 37.5 \omega_b \theta_\ast + 0.0942 \omega_b \Tilde{\mathcal{A}}_s + 0.281 \omega_b \tau  \\
    &+ 9.50 \omega_c \theta_\ast - 0.0073 \omega_c \Tilde{\mathcal{A}}_s  - 0.0511 \omega_c \tau  \\
    &- 476 \theta_\ast^{2} - 0.357 \theta_\ast \Tilde{\mathcal{A}}_s \\
    &+ 1.31 \theta_\ast n_s + 1.77 \theta_\ast \tau + 10.3 \theta_\ast \\
    &- 0.00076 \Tilde{\mathcal{A}}_s^{2} + 0.00073 \Tilde{\mathcal{A}}_s n_s 
    + 0.000162 \Tilde{\mathcal{A}}_s \tau \\
    &+ 0.00492 \Tilde{\mathcal{A}}_s  - 0.00378 n_s \tau  - 0.0364 \tau^{2} + 0.0137 \tau 
\end{split}
\end{equation}
where $\Tilde{\mathcal{A}}_s = \ln(10^{10} A_s)$ has been defined for convenience,
and the emulation for $H_3$ takes the form
\begin{equation}
\begin{split}
    H_3^{\text{(Z25)}} = & 
    - 82.7 \omega_b^{2} - 3.96 \omega_c^{2} + 0.298 n_s^{2}\\
    & - 18.9 \omega_b \omega_c - 6.52 \omega_b n_s  + 1.44 \omega_c n_s \\
    & + 6.26 \omega_b + 1.20 \omega_c - 0.0669 n_s -0.0375 \\
    & - 47.1 \omega_b \theta_\ast + 0.103 \omega_b \Tilde{\mathcal{A}}_s + 0.272 \omega_b \tau  \\
    &+ 39.7 \omega_c \theta_\ast 
    - 0.0998 \omega_c \Tilde{\mathcal{A}}_s - 0.0776 \omega_c \tau \\
    &- 1294 \theta_\ast^{2} + 2.23 \theta_\ast \Tilde{\mathcal{A}}_s + 3.13 \theta_\ast n_s \\
    &+ 0.192 \theta_\ast \tau + 16.1 \theta_\ast - 0.00331 \Tilde{\mathcal{A}}_s^{2} \\
    &- 0.00494 \Tilde{\mathcal{A}}_s n_s + 0.000483 \Tilde{\mathcal{A}}_s \tau + 0.00427 \Tilde{\mathcal{A}}_s  \\
    &- 0.00593 n_s \tau  - 0.0461 \tau^{2} + 0.00765 \tau 
\end{split}
\end{equation}
Symbolic representations for additional observables, as well as inverse mappings from observables to cosmological parameters, are provided in the accompanying {\sc MomentEmu} GitHub repository notebook. For brevity, these lengthy expressions are not reproduced here.

\balance

%%%%%%%%%%%%%%%%%%%%%%%%%%%%%%%%%%%%%%%%%%%%%%%%%%

% Don't change these lines
\bsp	% typesetting comment
\label{lastpage}
\end{document}